\documentstyle[preprint,eqsecnum,aps]{revtex}

\def\gsim{\mathrel{\scriptstyle{\buildrel > \over \sim}}}

\begin{document}
\newcommand{\be}{\begin{equation}}
\newcommand{\ee}{\end{equation}}
\newcommand{\bea}{\begin{eqnarray}}
\newcommand{\eea}{\end{eqnarray}}
\def\u{\uparrow}
\def\d{\downarrow}
\def\bi{{\bf i}}
\def\bj{{\bf j}}
\def\bl{{\bf l}}
\def\bk{{\bf k}}
\def\bx{{\bf x}}
\def\br{{\bf r}}
\def\bR{{\bf R}}
\title{Spin diffusion of the t-J model}

\author{  J. Bon\v ca$^{a,b}$ and J. Jakli\v c$^b$ }

\address{
$^{a}$Theoretical Division and Center for Nonlinear Studies,\\
Los Alamos National Laboratory, Los Alamos, New Mexico 87545,\\
$^b$J. Stefan Institute, University of Ljubljana, 61111 Ljubljana,
Slovenia.}
\maketitle

\begin{abstract}
  The spin-diffusion constant of the 2D $t-J$ model is calculated for
the first time using an analytical approach at high temperatures and a
recently-developed numerical method based on the Lanczos technique
combined with random sampling in the intermediate temperature
regime. A simple relation, $\sigma = D_s\chi$, between spin
conductivity and spin diffusion is established and used to calculate
the latter. In the high-temperature and low-doping limit the
calculated diffusion constant agrees with known results for the
Heisenberg model. At small hole doping, $D_s$ increases approximately
linearly with doping, which leads us to an important conclusion that
hopping processes enhance spin diffusion at high temperatures. At
modest hole doping, $\delta\sim 0.25$, diffusion exhibits a
nonmonotonic temperature dependence, which indicates anomalous spin
dynamics at small frequencies.
\end{abstract}


\newpage
\section{INTRODUCTION}

   There appears to be a rising interest in the spin dynamics of
high-temperature superconductor compounds. The research in this field is
mostly concentrated around the spin-lattice-relaxation rate, $1/T_1$.
In particular, recent measurements of $1/T_1$ in $\rm Sr$-doped $\rm
La_2CuO_4$ \cite{imai} display rather unusual temperature and doping
dependences. Sokol {\it et al.} \cite{sokol} have calculated $1/T_1$
for the undoped case and found good agreement with experiments. They
also calculated the spin-diffusion constant, which gives a
short-wavelength contribution to the nuclear-relaxation rate at high
temperatures. The relative strength of this diffusive contribution to
the relaxation rate depends strongly on the hyperfine structure, which
differs in the case of Cu and O nuclei \cite{gelfand,monien}. Spin
diffusion can serve also as a  general numerical test probe for
the spin dynamics of the system at small frequencies and momenta.

  The spin-diffusion constant of a paramagnet in the limit of high
temperature was first calculated by de Gennes using a moment expansion
of the spin-susceptibility function \cite{degennes}.  The method was
later improved by Bennett and Martin \cite{bennett} using an exact
spectral representation of the spin-correlation function in the
hydrodynamic limit. Morita \cite{morita} investigated a theoretical
criterion for the occurrence of spin-diffusion by introducing a
friction function. He also calculated diffusion constants for various
lattices and values of spin.

  In this paper we  present a finite-temperature calculation of
the spin-diffusion constant in the $t-J$ model on a square lattice
\be
H =  -t\sum_{<{\bf i,j}>,s}
\left(c_{\bi s}^\dagger c_{\bj s}^{ }  + {\rm h.c.}\right)
+ J\sum_{<\bi,\bj>}\left( {\bf S}_\bi\cdot{\bf S}_\bj -
n_\bi n_\bj /4\right),
\label{tj}
\ee
where ${<{\bf i,j}>}$ represents summation over pairs of nearest
neighbors, $c_{\bi,s}^\dagger$ and $c_{\bi,s}$ are creation and
annihilation fermion operators restricted to the basis of
single-fermion occupation number, and ${\bf S_\bi}$ is a spin-1/2
operator.

To obtain finite-temperature static and dynamic properties of the $t-J$
model, we use a recently developed numerical method, based on the Lanczos
exact-diagonalization technique combined with random sampling
\cite{prelovsek}. First we present an alternative method of calculating
spin diffusion, which is based on the spin-conductivity function.
Next, we present an analytical high-temperature result for $D_s$ in
two limits of the $t-J$ model: a) $t=0$ and b) $J=0$. In the case of
zero hole-doping, $\delta = 0$, our analytical method reproduces the
known result for $D_s$ of the Heisenberg model. Next, we briefly
discuss the finite-temperature Lanczos method, and then present
results for $D_s$ in the $t-J$ model at arbitrary parameter values.
Last, we present conclusions.

\section{The method}

\subsection{Spin-diffusion constant}

  We start by noting that the total $z$-component of the spin operator
$S^z=\sum_\br S^z(\br)$ is a constant of motion from which follows
the operator-continuity relation
\begin{equation}
{\partial S^z(\br,t)\over\partial t}+{\bf \nabla}\cdot\bj_s(\br,t)=0,
\label{cont}
\end{equation}
where $\bj_s$ is the spin-$z$-current operator.  Now, we suppose that
at finite temperatures the hydrodynamic description can be applied to
spin relaxation in the $t-J$ model in which the net flow of
magnetization from the region of large $S^z$ towards small $S^z$ is governed
by the diffusion equation
\begin{equation}
{\partial \langle S^z(\br,t)\rangle_{non-eq}\over\partial t}-
D_s \nabla^2\langle S^z(\br,t)\rangle_{non-eq}=0.
\label{diffeq}
\end{equation}
It is important to notice that contrary to eq.~(\ref{cont}), which is
an operator equation, the diffusion equation is valid only with
brackets $<>_{non-eq}$, which represent nonequilibrium ensemble
averages.

Applying linear response, hydrodynamics and the
fluctuation-dissipation theorem one can obtain the following form of
the Fourier transform of the spin-correlation function $S(\br,t)=
\langle S^z(\br,t) S^z(0,0)\rangle$ \cite{forster},
\bea
S(\bk,\omega)&=&
\langle S^z(\bk,\omega) S^z(-\bk,-\omega)\rangle = \nonumber \\
                  &\approx&{2\over 1-e^{-\beta\omega}}\times {\omega D_s
k^2\chi\over\omega^2+(D_s k^2)^2},
\label{skom}
\eea
valid for small $k$ and $\omega$.  The $S^z(\bk,\omega)$ is a spacial
and temporal Fourier transform of $S^z(\br,t)$. We use units in which
$\hbar=k_B=1$. In equation (\ref{skom}) $\beta$ is the inverse
temperature, $\chi$ is the static  magnetic susceptibility, and
$D_s$ is the spin diffusion constant. Note, that the last factor in
eq. (\ref{skom}) is the imaginary part of the dynamic susceptibility
function in the hydrodynamic limit. The brackets $<>$ in equation
(\ref{skom}) and all the following equations represent thermodynamic
equilibrium averages. According to Onsager, the spontaneous
equilibrium fluctuations of the spin, described by the
spin-correlation function $S(\bk,\omega)$, obey the same diffusion
equation as do the nonequilibrium - induced spin fluctuations in
eq. (\ref{diffeq}).

Next, we define the spin conductivity as
\bea
\sigma_x(\bk,\omega)&=&{1-e^{-\beta\omega}\over 2\omega}
\langle j_x(\bk,\omega)j_x(-\bk,\-\omega)\rangle =\nonumber \\
                       &=&{1-e^{-\beta\omega}\over 2\omega}
\int_{-\infty}^\infty dt\sum_{\br}e^{i(\bk\cdot\br-\omega t)}
\langle j_x(\br,t)j_x(0,0)\rangle,
\label{sigma}
\eea
where $j_x(\br,t)$ is the $x-$component of the spin current and
$j_x(\bk,\omega)$ its Fourier transform. The spin conductivity
$\sigma(\bk, \omega)$ can be related to the spin response function
$S(\bk, \omega)$ using the continuity relation (\ref{cont}) in
$(\bk,\omega)$ space: $-\omega S^z(\bk, \omega)
+\bk \cdot\bj(\bk, \omega)=0$:
\be
\sigma_x (\bk,\omega)={1\over 2}\omega(1-e^{-\beta\omega})
{S(\bk,\omega)\over k^2}.
\label{sigma1}
\ee
Combining eqs.\ (\ref{skom}) and (\ref{sigma1}) we derive the
following simple Einstein relation between the spin conductivity and
spin-diffusion constant:
\be
\sigma=\sigma_x(\bk=0,\omega=0) = D_s\chi.
\label{sigd}
\ee
\subsection{${\bf D_s}$ in the high temperature limit}

  In the high-temperature limit of the Heisenberg model it is
reasonable to expect that the spin conductivity, $\sigma(\omega)$ at
$\bk=0$, can be approximated with a Gaussian form \cite{bennett}. We
have checked this assumption numerically. However, this is in general
no longer true in the case of the $t-J$ model, predominantly due to
the coexistence of two different energy scales. There are two limiting
cases of the $t-J$ model, however, for which we expect that a Gaussian
form is still a reasonably good ansatz: a) $J=1,~t=0,
\delta\ll1$ and b) $J=0,~t=1,~\delta\ll1$, where $\delta$ represents
hole-doping. Note, that the above assumption must be limited to small
hole-doping regime. In the high-doping regime, the spin conductivity
has a Drude-like (Lorentzian) form.

While a direct computation of $\sigma(\omega=0)$ would be subject to
all the usual vagaries of numerical computation, we can avoid these
(after taking the assumption of the Gaussian form of
$\sigma(\omega)$) by taking a ratio (see (\ref{ds})) of the first two
nonzero moments of the spin conductivity:
\bea
\langle\omega^0\rangle&=&
\int_{-\infty}^\infty d\omega\ \sigma(\omega)=
{\pi\over T}\sum_{\br}\langle j_x(\br,0)j_x(0,0)\rangle,\nonumber \\
\langle\omega^2\rangle&=&
\int_{-\infty}^\infty d\omega\ \omega^2 \sigma(\omega)=
{\pi\over T}\sum_{\br}\langle
        ( i{\partial\over\partial t})j_x(\br,t)
        (-i{\partial\over\partial t})j_x(0  ,t))\rangle,
\label{mom}
\eea
where the time derivatives are to be taken at $t=0$ and the index
$\br$ runs over all lattice sites. The spin-current operator
$j_x(\br)$ for the $t-J$ model can be calculated using the continuity
relation (\ref{cont}) together with the relation
${\partial\over\partial t} S^z(\br,t)=-i[S^z(\br,t),H]$. Taking
into account the fact that all the operators are now defined on a
discrete lattice, we obtain
\be
j_x(\br)=-{i\over 2}
 \left[J\left(S^+_\br S^-_{\br +\hat \bx}-{\rm h.c.}\right)
      -t\left(c_{\br\u}^\dagger c_{\br +\hat \bx\u}^{ }
      - c_{\br\d}^\dagger c_{\br +\hat \bx\d}^{ } - {\rm h.c.}
\right)\right],
\label{j}
\ee
with $\hat \bx$ being the unit vector in the $x$-direction. To avoid
confusion we note again that $j_x$ represents the $x$-component of the
spin-$z$-current operator.  In order to calculate the first two
moments (\ref{mom}), one has to compute also the time derivative of
the spin current.  Taking the high-temperature expectation
values we arrive after a somewhat tedious but straightforward
calculation at the following expressions for the first two moments:
\bea
\langle\omega^0\rangle &=& \left({t^2\over 2}\delta\left(1-\delta\right) +
{J^2\over 8}\left(1-\delta\right)^2\right){\pi\over T},\nonumber\\
\langle\omega^2\rangle &=&
\left({t^4\over 2}\delta\left(1-\delta\right)^2\left(9+\delta\right) +
{J^4\over 16}\left(1-\delta\right)^2\left(5-3\delta\right) +
\hbox{\it O}(t^2J^2)\right){\pi\over T}.
\label{moments}
\eea
In the above equation, the second moment $\langle\omega^2\rangle$ is
correct only in the two limits of interest, {\it i.e.} at a) $J=0$
or b) $t=0$. Assuming a Gaussian functional form for the spin
conductivity, we can express the diffusion constant in eq. (\ref{sigd})
in terms of  the first two moments (\ref{mom})
\be
D_s = {1\over \sqrt{2\pi}}{1\over \chi}
\sqrt{\langle\omega^0\rangle^3\over \langle\omega^2\rangle},
\label{ds}
\ee
which leads to the final expressions for the spin-diffusion constant
for the two limiting cases:
\be
D_s = \left\{ \begin{array}{ll}
J{\left(1-\delta\right)\over 2}\sqrt{\pi\over
\left(5-3\delta\right)}&;\mbox{ for $t=0$},\\
t\delta
\sqrt{2\pi\over \left(9+\delta\right)\left(1-\delta\right)}&;\mbox{ for
$J=0.$}
\end{array}
\right.
\label{diff}
\ee
To test the method, we compare the diffusion constant (\ref{diff}) in the
Heisenberg limit $(t=0,~\delta=0)$ with the known result for the
Heisenberg model at high-temperatures $D_s/J=\sqrt{\pi\over 20}$
\cite{sokol,bennett} and find exact agreement.
At $t=0$ and small hole-doping $\delta$, $D_s$ decreases linearly with
doping as $D_s/J\sim \sqrt{\pi\over 20}(1-{7\over 10}\delta)$. This is
simply related to the fact that the number of spin-current carriers
(spins) decreases with doping. In the opposite limit, when $J=0$,
spin-diffusion (\ref{diff}) increases linearly with doping as
$D_s/t\sim (2\sqrt{10}/3)\sqrt{\pi\over 20}\delta$. It is
straightforward to check that at zero doping and $J=0$, $D_s/t=0$,
since the spin-current (\ref{j}) at $\delta=0$ is zero. The linear
increase of $D_s$ with doping $\delta$ can in this case be explained
by the linear increase of spin-current carriers
(holes). Interestingly, the coefficient of the linear increase of
$D_s$ in the $J=0$ limit is larger than the coefficient of decrease in
the $t=0$ limit. Since the physical parameter region (relevant for
cuprates), where $J/t\sim 0.3$, is not accessible by the present
analytic method, we make the following assumption: neglecting the
combined hopping and spin-flip contributions to the diffusion constant
we predict that in the high-temperature regime and at $J/t=0.3$, $D_s$
increases linearly with doping.

\subsection{Numerical method}

Having exhausted the possibilities of analytical calculation we now
turn to the numerical calculation of $D_s$ in physically relevant
parameter regimes. We used a finite-temperature Lanczos method,
recently developed by one of the present authors (J.J.) in cooperation
with P. Prelov\v sek.  Since the details of the method have been
published \cite{prelovsek}, we only briefly summarize the main steps
of the method.

Taking into account the translational symmetry of the system, the spin
conductivity (\ref{sigma}) at $\bk=0$ can be written as
\bea
\sigma(\omega)&=&{1-e^{-\beta\omega}\over 2N\omega}
\int_{-\infty}^\infty dt~e^{-i\omega t}
\langle J_x(t)J_x(0)\rangle,\label{sigma0}\\
J_x(t)&=&\sum_{\br}j_x(\br,t),\nonumber
	\eea
where $N$ is the number of sites. According to the finite-temperature
Lanczos method \cite{prelovsek}, the thermodynamic equilibrium average
in eq. (\ref{sigma0}) is expressed in terms of functions
$\vert\psi_m^n\rangle$, $\vert\tilde \psi_k^n\rangle$ and corresponding
energies $\epsilon_{nm}$, $\tilde\epsilon_{nk}$, generated by the
Lanczos method on a finite cluster with $N$ sites,
\bea
\langle J_x(t)J_x(0)\rangle &=& Z^{-1}\sum_n^{N_0}\sum_{m,k}^M
\langle n\vert \psi_m^n\rangle e^{-\beta \epsilon _{nm}}
e^{i(\epsilon_{nm}-\tilde \epsilon_{nk})t}\langle\psi_m^n\vert
J_x\vert\tilde \psi_k^n\rangle
\langle\tilde \psi_k^n\vert J_x\vert n\rangle,\label{siglanc}\\
Z &=& \sum_n^{N_0}\sum_m^M \vert\langle n\vert \psi_m^n\rangle\vert ^2
e^{-\beta \epsilon_{nm}}\nonumber,
\eea
where states $\vert n\rangle$ for $n=1$ to $N_{st}$ represent a
complete orthonormal basis set of the Hamiltonian H. In the first
summation, only $N_0\ll N_{st}$ of states $\vert n\rangle$ are randomly
chosen for partial random sampling of basis states. The functions
$\vert\psi_m^n\rangle$ and energies $\epsilon_{nm}$ are obtained as
eigenvectors and eigenvalues of the tridiagonal form of the
Hamiltonian, generated by the Lanczos procedure,
\bea
H\vert\phi_m^n\rangle &=& b_{nm}\vert\phi_{m-1}^n\rangle
+ a_{nm}\vert\phi_m^n\rangle + b_{n(m+1)}\vert\phi_{m+1}^n\rangle
\label{lanc}\\
m&=&0,\ldots,M\ll N_{st};~b_{n0}=b_{n(M+1)}=0,\nonumber
\eea
starting from the initial wavefunction $\vert \phi_0^n\rangle = \vert
n\rangle$. Similarly, $\vert\tilde\psi_k^n\rangle$ and $\tilde
\epsilon_{nk}$ are generated by the Lanczos procedure, analogous to
(\ref{lanc}), but with a different initial wavefunction
$\vert\tilde\phi_0^n\rangle = J_x\vert n\rangle /\sqrt{\langle n\vert
J_x^2\vert n\rangle}$.

The static magnetic susceptibility $\chi=\beta S(\bk=0)$ is evaluated with
the same method, adapted to the calculation of static quantities.
Here, the equal-time spin correlation function is expressed as
\bea
S(\bk)=Z^{-1}\sum_n^{N_0}\sum_m^Me^{-\beta\epsilon_{nm}}
\langle n|\psi_m^n\rangle\langle \psi_m^n |S^z(\bk) S^z(-\bk)|n\rangle,
\eea
where $S^z(\bk)=N^{-1/2}\sum_\br e^{-i\bk\cdot\br}S^z(\br)$,
and the rest of notation is the same as in Eq.~(\ref{siglanc}).

The method yields very accurate results even for a severely reduced
number of Lanczos steps $M\ll N_{st}$ and number of random states
$N_0\ll N_{st}$.  Provided that $N_0$ is large enough,
$\sigma_x(\omega)$ has correct frequency moments
$\langle\omega^p\rangle$ in the limit $T\to\infty$ for $p=0,...,M$
(for a finite cluster). Similarly, at finite temperatures the
corresponding double series for the moments of $\sigma(\omega)$ are
correct up to $\beta^{p+1}\langle\omega^{M-p}\rangle,$ for
$p=0,\ldots,M$. The method also gives reliable results at $T=0$,
assuming that the number of Lanczos steps $M$ is larger than the
number of Lanczos steps required to obtain the corresponding
ground-state results. For a typical run, $M=100-150$ Lanzos steps and
sampling over $N_0=200-300$ were sufficient to obtain reasonably
accurate (within 5\%) results for spin conductivity and static spin
susceptibility.

\section{Numerical Results  and Conclusions}

  Using the finite-temperature Lanczos method we computed the
spin-diffusion constant of the $t-J$ model in the physically relevant
parameter regime, {\it i.e.} at $J/t=0.3$ and for four different
hole-dopings, $\delta=0,~1/16,~1/8$ and $1/4$.  The calculation was
performed on a $4\times 4$ cluster with periodic boundary conditions.

  In Fig.~1 we present numerical results for the renormalized
spin-diffusion constant $D_s/J$ of the Heisenberg model as a function of
$T/J$. For comparison we present on the same graph  $D_s/J$ obtained by
the high-temperature expansion by Sokol {\it et al.}, $D_s/J =
\sqrt{\pi\over 20}(1+{21\over 20}\beta)$, with $\beta = J/T$. Numerical
results agree well with the high-temperature predictions down to
temperatures $T/J\sim 1.5$, where $D_s$ obtained by the numerical
method saturates. In the $T\to 0$ limit, $D_s$ has to diverge, since
the Heisenberg model at $T=0$ has finite, nonzero spin stiffness
\cite{schulz}.  Saturation of $D_s$ is therefore a finite-size
effect. Nevertheless, the agreement with the high-temperature
expansion at higher temperatures is quite encouraging. In  contrast
to the finite-size $T=0$ calculations, where results are strongly
size-dependent, the finite-temperature Lanczos method even on a small
cluster size provides results which are correct in the thermodynamic
limit. We expect results to be size independent as long as the
temperature is larger than the average discrete level spacing
$\epsilon$, which is a consequence of the finite-size system. At
finite doping, where the low-lying energy sector becomes more dense,
we expect size-independent results down to much smaller temperatures.

  We now turn to a doped case. In Fig.~2 we plot $D_s/J$ as a function
of temperature for four different values of doping at $J/t=0.3$. As
predicted by the infinite-temperature moment expansion, $D_s/J$ is at
high temperatures, $T\gsim t$, increasing approximately linearly with
doping $\delta$ for $\delta \ll 1$. In contrast to the undoped case
where $D_s$ as a consequence of finite-size effects saturates around
$T\sim J$, $D_s$ at finite doping diverges with decreasing temperature
(even in the finite cluster). This is an indication of the formation
of a quantum coherent state, which has a finite spin-stiffness at
$T=0$. $D_s$ at small doping, $\delta\leq 1/8$, displays a monotonic
$1/T$-like temperature dependence, similar to the undoped
case. Somewhat unexpected is the nonmonotonic temperature dependence
of $D_s$ at moderate doping, $\delta=1/4$. At high temperatures,
$D_s/J$ decreases with falling temperature, reaches a minimum around
$T\sim 1.5~J$, and when the quantum coherent regime sets in at even
lower temperatures, sharply increases. We attribute this unusual
nonmonotonic temperature dependence of the diffusion constant to the
increase with doping of the number of low-energy short-lived many-body
states. An enhanced density of low-energy states contributes to
enhanced inelastic scattering, which consequently leads to a decrease
of the diffusion constant. This picture is in agreement with the
enhanced entropy at low temperatures \cite{puttika} and strong
charge-carrier scattering, which is manifested in the linear
temperature dependence of the resistivity in the $t-J$ model
\cite{jaklic}.

  In conclusion, we have calculated the spin-diffusion constant $D_s$
for the $t-J$ model. In the high-temperature limit, $D_s$ can be
approximately calculated via the moment expansion of the
spin-conductivity function at $t=0$ or $J=0$ and for small doping
$\delta$. In the Heisenberg limit, this method gives the correct
result for $D_s$. At small doping, $\delta\ll 1$, and at $t=0$ (the
dilute Heisenberg model), $D_s/J$ decreases linearly with doping,
indicating that the spin-current carriers are localized spins.  At
$J=0$ and $\delta\ll 1$ $D_s/t$ is increasing linearly with $\delta$
which  is consistent with the fact that the spin-current carriers in
this limit are mobile holes.

  At finite temperatures, we used a recently developed Lanczos method
to obtain $D_s$ as a function of temperature and doping. We found good
agreement of $D_s$ in the Heisenberg limit with the high-temperature
expansion result at higher temperatures.  In agreement with our
analytical predictions, $D_s$ is increasing approximately linearly
with doping at $J/t\sim 0.3$ in the high-temperature and low-doping
regime. Since the diffusion constant decreases with hole doping in a
dilute Heisenberg model, {\it i.e.} at $t=0$, we can conclude that
hopping processes contribute substantially to spin diffusion at high
temperatures. At moderate doping, $\delta = 1/4$, and high
temperature, $T\gsim J$, $D_s$ decreases with decreasing
temperature. This anomalous behavior can be associated with the
recently reported anomalous spin dynamics of the doped $t-J$ model at
low-frequencies \cite{jaklic1} and enhanced entropy at low
temperatures
\cite{puttika}.

\section{Acknowledgments}

  We gratefully acknowledge valuable discussion with T.M.\ Rice and
P. Prelov\v sek.  We are grateful to D. Rabson and J.P. Rodrigues
for several illuminating discussions and critically reading the
manuscript.

\newpage
\begin{figure}
\caption[]{ Spin-diffusion constant $D_s$ of the Heisenberg model
vs.\ temperature obtained by the finite-temperature Lanczos method
(full circles). For comparison we present the
high-temperature-expansion results obtained by Sokol {\it et al.}
{\cite{sokol}} (dashed line).\label{f:1}}
\end{figure}

\begin{figure}
\caption[]{ Spin-diffusion constant $D_s$ of the $t-J$ model vs.\
 $T/t$ at $J/t=0.3$ and various doping values. The arrow above the
 temperature axis points to $T=J$. The accuracy of results at finite
 doping is 5\%.\label{f:2}}
\end{figure}


\begin{references}

\bibitem{imai} T. Imai, C.P. Slichter, K. Yoshimura, and K. Kosuge,
  Phys. Rev. Lett. {\bf 70}, 1002 (1993).

\bibitem{sokol} A. Sokol, E. Gagliano, and S. Bacci,
 Phys. Rev. Lett. {\bf 47}, 14646 (1993).

\bibitem{gelfand} S. Chakravarty {\it et al.,} Phys. Rev. B {\bf 43},
 2796 (1991).


\bibitem{monien} A.J. Millis and H. Monien, preprint.

\bibitem{degennes}P.G. De Gennes, J. Phys. Chem. Solids {\bf 4}, 223
 (1958).

\bibitem{bennett} H.S. Bennett and P. C. Martin, Phys. Rev. {\bf 138},
 A 608 (1965).

\bibitem{morita} T. Morita, Phys. Rev. B {\bf 6}, 3385 (1972).

\bibitem{prelovsek} J. Jakli\v c and P. Prelov\v sek, Phys. Rev. B {\bf
 49}, 5065 (1994).

\bibitem{forster} Dieter Forster, {\it Hydrodynamic Fluctuations,
  Broken Symmetry, and Correlation Functions}, Frontiers in Physics
  vol. {\bf 47}, (1975).

\bibitem{schulz} R.R.P. Singh and D.A. Huse, Phys. Rev. B
 {\bf 40}, 7247 (1989); J.I. Igarashi, Phys. Rev. B {\bf 46}, 10763
 (1992); T. Einarsson and H.J. Schulz, preprint.


\bibitem{puttika} B. Puttika, unpublished.

\bibitem{jaklic} J. Jakli\v c and P. Prelov\v sek, Phys. Rev. B {\bf
 50}, 7129 (1994).

\bibitem{jaklic1} J. Jakli\v c and P. Prelov\v sek, preprint
IJS-TP-94/24, SISSA, 9410085.


\end{references}
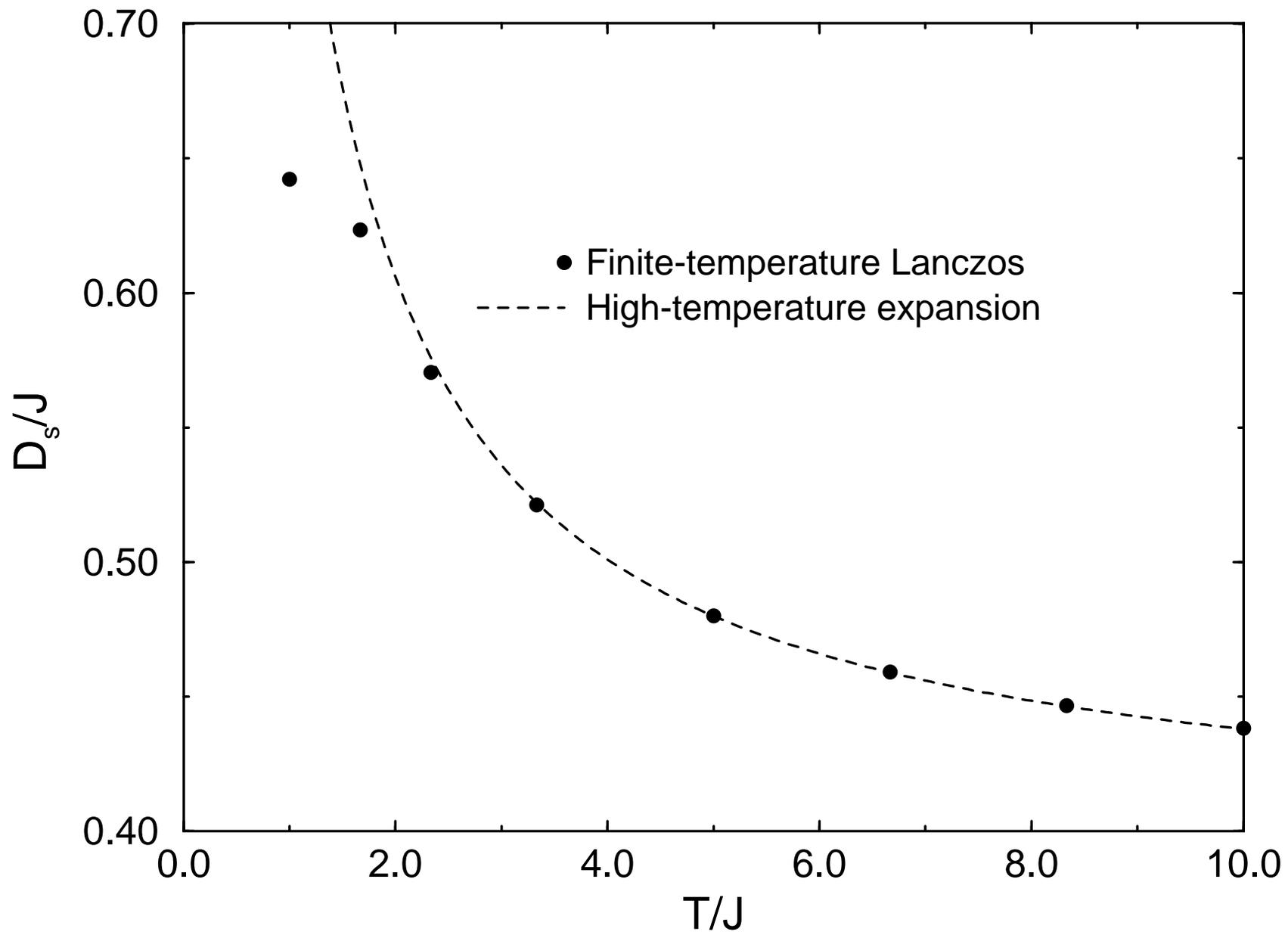
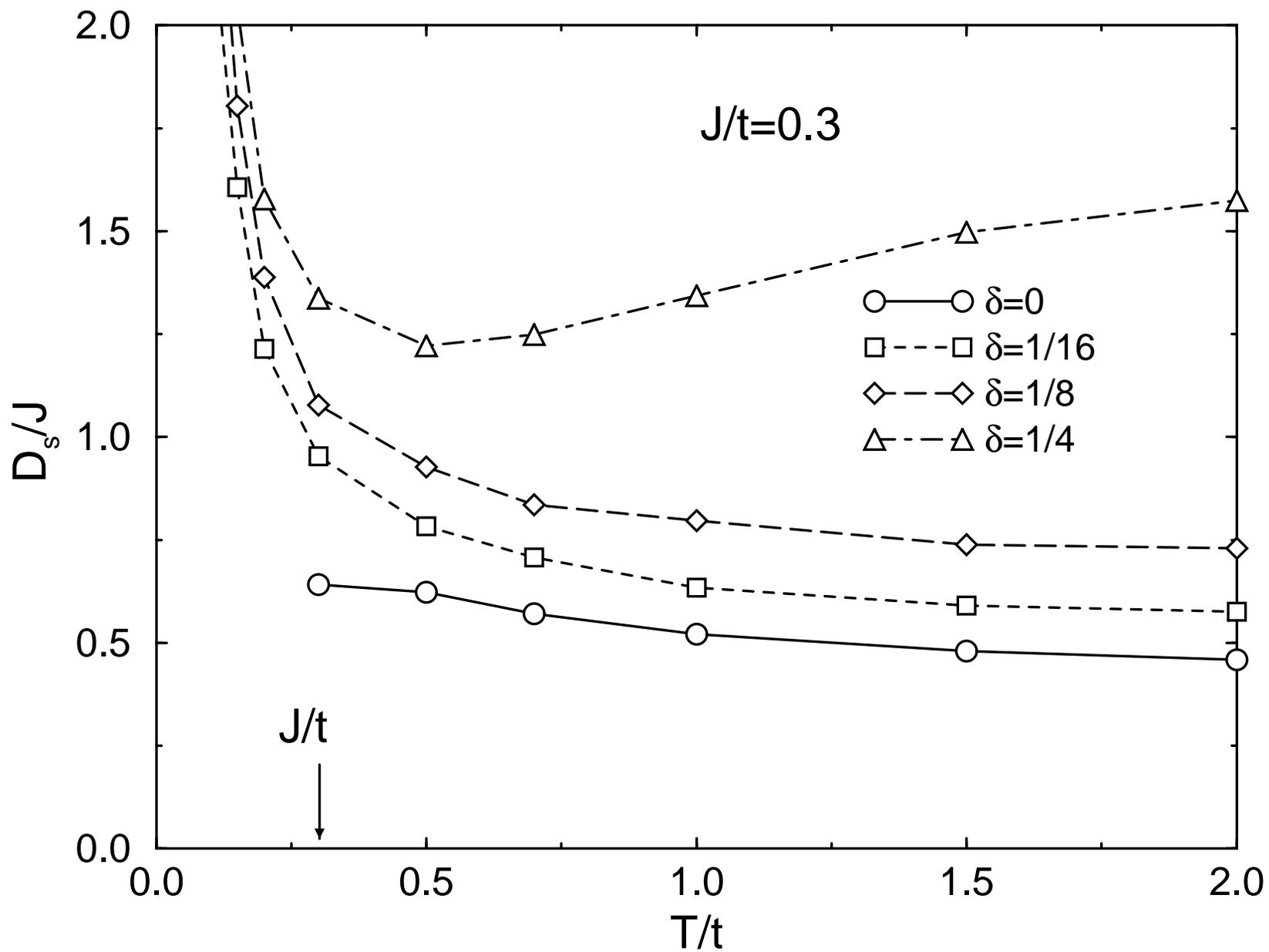
\end{document}